\documentclass[aps,pra,twocolumn,showpacs,superscriptaddress]{revtex4}

\usepackage{amsfonts,mathrsfs,amsmath,amsthm,amssymb}
\usepackage{bm}
\usepackage{graphicx,epsfig}
\usepackage[colorlinks=true,breaklinks=true,linkcolor=blue,citecolor=blue]{hyperref}

\newcommand{\be}{\begin{eqnarray}}
\newcommand{\ee}{\end{eqnarray}}

\makeatletter
    
    \newcommand{\Rmnum}[1]{\expandafter\@slowromancap\romannumeral #1@}
\makeatother

\begin{document}

\title{Unambiguous discrimination between two unknown qudit states}

\author{Tao Zhou}
\affiliation{Department of Physics, Tsinghua University, Beijing, 100084, China}
\affiliation{Department of Physics, Sichuan University, Chengdu, 610064, China}

\date{\today}


\begin{abstract}
We consider the unambiguous discrimination between two unknown 
qudit states in $n$-dimensional ($n\geqslant2$) Hilbert space. By
equivalence of unknown pure states to known mixed states and with
the Jordan-basis method, we demonstrate that the optimal success
probability of the discrimination between two unknown states is
independent of the dimension $n$. We also give a scheme for a
physical implementation of the programmable state discriminator that can
unambiguously discriminate between two unknown states with optimal
probability of success.
\end{abstract}
\maketitle

\section{Introduction}\label{sec1}
As a recent development, the possibility of unambiguous
discrimination between unknown quantum states can be potentially
useful for many application in quantum communication and quantum
computing. A universal device that can unambiguously discriminate
between two unknown states has been constructed by Bergou and
Hillery \cite{r1}. It has three registers, labeled A, B and C, and
each register can store a qubit that is in some arbitrary state. In
their work, it is assumed that register A is prepared in the state
$|\psi_1\rangle$, register C is prepared in the states
$|\psi_2\rangle$, and register B is guaranteed to be prepared in
either $|\psi_1\rangle$ or $|\psi_2\rangle$. Here, $|\psi_1\rangle$
and $|\psi_2\rangle$ are the states to be distinguished, which are
both unknown
\begin{equation}
|\psi_1\rangle=a|0\rangle+b|1\rangle,\quad
|\psi_2\rangle=c|0\rangle+d|1\rangle,
\end{equation}
where all the parameters $a$,$b$,$c$ and $d$ are arbitrary unknown
complex variables satisfying the normalization equation
$|a|^2+|b|^2=1$ and $|c|^2+|d|^2=1$. Furthermore, it is assumed that
register B is prepared in the state $|\psi_1\rangle$ with
probability $\eta_1$ and in the state $|\psi_2\rangle$ with
probability $\eta_2$, such that $\eta_1+\eta_2=1$, which guarantees
that the state in register B is always one of these two states. The
states viewed as a program which are sent into registers A and C
(called program registers) are called program states, while the
unknown state to be conformed which is sent into the register B
(called data register) is called data state. The device constructed
here can measure the total input states
\begin{eqnarray}
|\Psi_1\rangle
&=&|\psi_1\rangle_A|\psi_1\rangle_B|\psi_2\rangle_C,\nonumber\\
|\Psi_2\rangle &=&|\psi_1\rangle_A|\psi_2\rangle_B|\psi_2\rangle_C,
\end{eqnarray}
which are prepared with apriori probabilities $\eta_1$ and $\eta_2$.
With the symmetry properties of the input states, this device will
then, with some probabilities of success, tell us whether the
unknown state in the data register B matches the state stored in
program register A or C.

In later works, a series of new devices have been introduced and widely discussed by
other authors \cite{r2,r3,r4,r5,Sentis,PLA356.306}. In these schemes we may have $n_A$
and $n_C$ copies of the states in the program register A and C
respectively, and $n_B$ copies of the states in the data register B,
then our task is to distinguish the two states
\begin{eqnarray}
|\Psi_1\rangle &=& |\psi_1\rangle_A^{\otimes
n_A}|\psi_1\rangle_B^{\otimes n_B}|\psi_2\rangle_C^{\otimes n_C},\nonumber\\
|\Psi_2\rangle &=& |\psi_1\rangle_A^{\otimes
n_A}|\psi_2\rangle_B^{\otimes n_B}|\psi_2\rangle_C^{\otimes n_C},
\end{eqnarray}
with the minimum error or with the minimum inconclusive probability,
if the minimum error strategy or optimum unambiguous strategy is
applied, respectively \cite{r6}.

In both previous papers \cite{r2} and \cite{r6}, it has been proved
that the optimal success probability of discrimination between two
unknown qubits is an increasing function of $n_B$, the number of
copies in data register B. In this paper, we study the qudit
states in n-dimensional Hilbert space and demonstrate that the
optimal success probability of discrimination between two unknown
states is independent of the dimension $n$. Unlike the
discrimination between two known states, we cannot consider the
subspace spanned by the two states only, but we should consider the
full $n$-dimensional space, as the two states are completely unknown
to us. We adopt the equivalence between the discrimination of
unknown pure states and that of known mixed states, and then reduce
the problem to the unambiguous discrimination between two pure
states with Jordan-basis method. Finally, we get the optimal success
probability of the unambiguous discrimination between the two mixed
states and give the detection operators. The detection operators are
also applicable to the discrimination between the pure states
without averaging over them.

The organization of the paper is as follows. In Sec. \ref{sec2}, we
give a brief description of the programmable states discrimination,
and adopt the equivalence between the discrimination of unknown
states to that of known mixed states. In Sec. \ref{sec3}, we
introduced the Jordan basis for the mean input states. In Sec.
\ref{sec4}, we find the optimal unambiguous discrimination and
introduce its implementation. Finally, a brief summary is given in
Sec. \ref{sec5}.

\section{The programmable discrimination}\label{sec2}
In previous works, states such as
$|\psi\rangle=\cos(\phi/2)|0\rangle+\sin(\phi/2)|1\rangle$ have been
considered \cite{r7}. These states lie in a two-dimensional Hilbert
space. Here we introduce a more generalized state
\begin{equation}
|\psi\rangle=\sum_{i=1}^na_i|i\rangle,
\end{equation}
where $a_i$ is an arbitrary unknown complex variable. All these
states span a n-dimensional Hilbert space $\mathcal{H}$, for which
$|i\rangle$ form a mutually orthogonal basis. The two states we want
to distinguish are denoted by $|\psi_1\rangle$ and $|\psi_2\rangle$.
As we mentioned before, a copy of each of the two unknown states is
provided for the program registers A and C, denoted as
$|\psi_1\rangle_A$ and $|\psi_2\rangle_C$, respectively. The state
to be conformed is provided for the data register B as input. We
assume that the state in the data register B is guaranteed to be
prepared in $|\psi_1\rangle$ with probability $\eta_1$ and in
$|\psi_2\rangle$ with probability $\eta_2=1-\eta_1$ respectively.
Thus we have two possible inputs
\begin{eqnarray}
|\Psi_1\rangle
&=&|\psi_1\rangle_A|\psi_1\rangle_B|\psi_2\rangle_C,\nonumber\\
|\Psi_2\rangle &=&|\psi_1\rangle_A|\psi_2\rangle_B|\psi_2\rangle_C.
\end{eqnarray}
Here, $|\psi_1\rangle$ and $|\psi_2\rangle$ are completely arbitrary states
\begin{eqnarray}
|\psi_1\rangle &=& \sum_{i=1}^na_i|i\rangle,\nonumber\\
|\psi_2\rangle &=& \sum_{i=1}^nb_i|i\rangle,
\end{eqnarray}
that remain unknown throughout the entire discrimination process.

Since the inputs $|\Psi_1\rangle$ and $|\Psi_2\rangle$ are unknown,
and the states $|\psi_1\rangle$ and $|\psi_2\rangle$ making up these
two inputs can change from preparation to preparation, it is only
the pattern, middle states (B) matching the first (A) or the last
(C) state that is preserved from one preparation to the next.
Therefore, we can introduce the corresponding density operators
\begin{eqnarray}
\rho_1 &=& \{|\Psi_1\rangle\langle\Psi_1|\}_{av},\nonumber\\
\rho_2 &=& \{|\Psi_2\rangle\langle\Psi_2|\}_{av},
\end{eqnarray}
where the average is taken over the entire parameter space of states
$|\psi_1\rangle$ and $|\psi_2\rangle$. Here, we should notice that
the optimal strategy for discrimination between the two states is
the strategy that is optimal on average. That's to say, we can
unambiguously discriminate between $|\psi_1\rangle$ and
$|\psi_2\rangle$ as soon as we can unambiguously discriminate
between $\rho_1$ and $\rho_2$.

To this end, we now define the space and operators that we will
need. Let $\mathcal{H}$ be the n-dimensional space for the unknown
states and then the full space is $\otimes^3\mathcal{H}$. The space
of symmetric states in $\mathcal{H}\otimes\mathcal{H}$ is denoted by
$\Sigma$, which is an $n(n+1)/2$-dimensional subspace.
$|\Psi_1\rangle$ is an element of $S_1=\Sigma\otimes\mathcal{H}$ and
$|\Psi_2\rangle$ is an element of $S_2=\mathcal{H}\otimes\Sigma$.
Therefore, $S_0=S_1\cap S_2$, the intersection of $S_1$ and $S_2$,
is the space of symmetric states in the full space
$\otimes^3\mathcal{H}$. $S_0$ is a subspace of dimension
$n(n+1)(n+2)/6$, while $S_1$ and $S_2$ both $n^2(n+1)/2$. Let $S_3$
be the subspace of $\otimes^3\mathcal{H}$ generated by $S_1$ and
$S_2$, and the dimension of $S_3$ is $n(n+1)(5n-2)/6$. Let $S_4$ be
the orthogonal complement of $S_0$ in $S_1$, let $S_5$ be the
orthogonal complement of $S_0$ in $S_2$, and let $S_6$ be the
orthogonal complement of $S_0$ in $S_3$.

Clearly, the average in $\rho_1$ uniformly fills the symmetric
subspace of A and B and the entire subspace of C, whereas the
average in $\rho_2$ uniformly fills the symmetric subspace of B and
C and the entire subspace of A. Therefore, the corresponding density
operators averaged over the unknown states, can be expressed as
\begin{eqnarray}
\label{eq1}
\rho_1 &=& \frac{2}{n(n+1)}P_\Sigma\otimes\frac{1}{n}I_{C}\nonumber\\
&=& \frac{2}{n^2(n+1)}P_\Sigma\otimes I_{C},\nonumber\\
\rho_2 &=& \frac{1}{n}I_{A}\otimes\frac{2}{n(n+1)}P_\Sigma\nonumber\\
&=& \frac{2}{n^2(n+1)}I_{A}\otimes P_\Sigma,
\end{eqnarray}
where $P_\Sigma$ is the projection onto $\Sigma$, and $I$ onto
$\mathcal{H}$. $P_\Sigma$ can be expressed as
\begin{equation}
P_\Sigma=\sum_{i\leqslant j=1}^n |u_{ij}^{(2)}\rangle\langle
u_{ij}^{(2)}|,
\end{equation}
where $|u_{ij}^{(2)}\rangle\ (i\leqslant j=1,...,n)$ are the unique
unit vectors in the symmetric subspace $\Sigma$,
\begin{eqnarray}
|u_{11}^{(2)}\rangle&=&|11\rangle,\nonumber\\
|u_{12}^{(2)}\rangle&=&\frac{1}{\sqrt{2}}(|12\rangle+|21\rangle),\nonumber\\
&\vdots&\nonumber\\
|u_{nn}^{(2)}\rangle&=&|nn\rangle.
\end{eqnarray}

Consequently, we reduced the problem to discriminating between the
$[n^2(n+1)/2]$-dimensional spaces $S_1$ and $S_2$ in $S_3$, which is
equivalent to discriminating between subspaces $S_4$ and $S_5$ in
the $[2n(n+1)(n-1)/3]$-dimensional space $S_6$.

\section{Jordan basis}\label{sec3}
It has been shown in \cite{r8} that two nonorthogonal subspaces
$\mathcal{H}_1$ and $\mathcal{H}_2$ of a Hilbert space can be
unambiguously discriminated if we can find their canonical or Jordan
bases \cite{r9}. The definition of the Jordan bases is as follows.
The set of orthogonal and normalized basis vectors
$\{|g_1\rangle,...|g_n\rangle\}$ in $\mathcal{H}_1$ and
$\{|h_1\rangle,...,|h_n\rangle\}$ in $\mathcal{H}_2$ form Jordan
bases if and only if
\begin{equation}
\langle g_i|h_j\rangle=\delta_{ij}\cos\theta_i,
\end{equation}
where $\theta_i$ are the Jordan angles
$(\theta_1\leqslant\theta_2\leqslant...\leqslant\theta_n)$. In the
Jordan form, the full Hilbert space is decomposed into $n$
orthogonal subspaces. Here, we denote the $i$th subspace spanned by
$|g_i\rangle$ and $|h_i\rangle$ by $T_i$. Clearly, if the dimension
of $T_i$ is 1, we can get $\cos\theta_i=1$, and no further
discrimination is possible in $T_i$. On the other hand, if
$\cos\theta_i<1$, $|g_i\rangle$ and $|h_i\rangle$ are linearly
independent, and they can be distinguished as two known pure states,
which is familiar to us. Thus our task is reduced to some separate
discriminations in each subspace, between two known pure states.

Now, we will choose some bases to construct Jordan bases for the
density operators $\rho_1$ and $\rho_2$. First, we should notice
that the fully symmetric subspace of the three states $S_0$ must be
common to both inputs. Here, we denote the basis for the subspace
$S_0$ by $|u_{ijk}^{(3)}\rangle$, where $1\leqslant i\leqslant
j\leqslant k\leqslant n$ is satisfied, and there are $n(n+1)(n+2)/6$
these unique unit vectors, which can be expressed as follows,
\begin{widetext}
\begin{equation}
|u_{ijk}^{(3)}\rangle=\left\{ \begin{array}{ll} |ijk\rangle &
\textrm{if $i=j=k$}\\
\\
\frac{\displaystyle |ijk\rangle+|ikj|\rangle+|kij\rangle}{\displaystyle\sqrt{3}} & \textrm{if $i=j\neq k$}\\
\\
\frac{\displaystyle
|ijk\rangle+|jik\rangle+|jki\rangle}{\displaystyle\sqrt{3}} &
\textrm{if $i\neq j=k$}\\
\\
\frac{\displaystyle
|ijk\rangle+|jik\rangle+|ikj\rangle+|kij\rangle+|jki\rangle+|kji\rangle}{\displaystyle\sqrt{6}}
& \textrm{if $i\neq j \neq k$}
\end{array} \right..
\end{equation}
\end{widetext}
Then, the structure of the two density operators in Eq. (\ref{eq1}),
in particular the decomposition on the right-hand side, suggests
that we consider $|u_{ij}^{(2)}\rangle\otimes|\alpha\rangle$ and
$|\alpha\rangle\otimes|u_{ij}^{(2)}\rangle$, where $\alpha=1,...,n$
and $1\leqslant i\leqslant j\leqslant n$. These vectors form
orthogonal bases for $S_1$ and $S_2$, respectively. Thus the unique
unit vector $|u_{ijk}^{(3)}\rangle$ in $S_0$ can be expressed in
terms of either $S_1$ or $S_2$, since it is in both spaces. Here, we
neglect the vectors $|u_{ijk}^{(3)}\rangle$ that satisfy $i=j=k$,
and a direct calculation shows that
\begin{widetext}
\begin{equation}
|u_{ijk}^{(3)}\rangle=\left\{ \begin{array}{ll}
\sqrt{\frac{\displaystyle 2}{\displaystyle 3}}|u_{ik}^{(2)}\rangle|j\rangle+\sqrt{\frac{\displaystyle 1}{\displaystyle 3}}|ijk\rangle &
\textrm{if $i=j\neq k$}\\
\\
\sqrt{\frac{\displaystyle 2}{\displaystyle
3}}|u_{ij}^{(2)}\rangle|k\rangle+\sqrt{\frac{\displaystyle
1}{\displaystyle 3}}|jki\rangle & \textrm{if $i\neq j=k$}\\
\\
\frac{\displaystyle
1}{\displaystyle\sqrt{3}}(|u_{ij}^{(2)}\rangle|k\rangle+|u_{ik}^{(2)}\rangle|j\rangle+|u_{jk}^{(2)}\rangle|i\rangle)
& \textrm{if $i\neq j \neq k$}
\end{array} \right.,
\end{equation}
and
\begin{equation}
|u_{ijk}^{(3)}\rangle=\left\{ \begin{array}{ll}
\sqrt{\frac{\displaystyle 2}{\displaystyle
3}}|i\rangle|u_{jk}^{(2)}\rangle+\sqrt{\frac{\displaystyle
1}{\displaystyle 3}}|kij\rangle &
\textrm{if $i=j\neq k$}\\
\\
\sqrt{\frac{\displaystyle 2}{\displaystyle
3}}|j\rangle|u_{ik}^{(2)}\rangle+\sqrt{\frac{\displaystyle
1}{\displaystyle 3}}|ijk\rangle & \textrm{if $i\neq j=k$}\\
\\
\frac{\displaystyle
1}{\displaystyle\sqrt{3}}(|k\rangle|u_{ij}^{(2)}\rangle+|j\rangle|u_{ik}^{(2)}\rangle+|i\rangle|u_{jk}^{(2)}\rangle)
& \textrm{if $i\neq j \neq k$}
\end{array} \right..
\end{equation}
\end{widetext}

We can now introduce the vectors
\begin{eqnarray}
|g_{ijk}\rangle=\sqrt{\frac{\displaystyle 1}{\displaystyle
3}}|u_{ik}^{(2)}\rangle|j\rangle-\sqrt{\frac{\displaystyle
2}{\displaystyle 3}}|ijk\rangle\\
|h_{ijk}\rangle=\sqrt{\frac{\displaystyle 1}{\displaystyle
3}}|i\rangle|u_{jk}^{(2)}\rangle-\sqrt{\frac{\displaystyle
2}{\displaystyle 3}}|kij\rangle
\end{eqnarray}
for $i=j\neq k$,
\begin{eqnarray}
|g_{ijk}\rangle=\sqrt{\frac{\displaystyle 1}{\displaystyle
3}}|u_{ij}^{(2)}\rangle|k\rangle-\sqrt{\frac{\displaystyle
2}{\displaystyle 3}}|jki\rangle\\
|h_{ijk}\rangle=\sqrt{\frac{\displaystyle 1}{\displaystyle
3}}|j\rangle|u_{ik}^{(2)}\rangle-\sqrt{\frac{\displaystyle
2}{\displaystyle 3}}|ijk\rangle
\end{eqnarray}
for $i\neq j=k$,
\begin{widetext}
\begin{eqnarray}
|g_{ijk}\rangle &=&\frac{\displaystyle 3-\sqrt{3}}{\displaystyle
6}|u_{ij}^{(2)}\rangle|k\rangle-\frac{\displaystyle
3+\sqrt{3}}{\displaystyle
6}|u_{ik}^{(2)}\rangle|j\rangle+\frac{\displaystyle
\sqrt{3}}{\displaystyle 3}|u_{jk}^{(2)}\rangle|i\rangle\nonumber\\
|g_{ijk}^{'}\rangle &=&\frac{\displaystyle 3-\sqrt{3}}{\displaystyle
6}|u_{ik}^{(2)}\rangle|j\rangle-\frac{\displaystyle
3+\sqrt{3}}{\displaystyle
6}|u_{ij}^{(2)}\rangle|k\rangle+\frac{\displaystyle
\sqrt{3}}{\displaystyle 3}|u_{jk}^{(2)}\rangle|i\rangle
\end{eqnarray}
and
\begin{eqnarray}
|h_{ijk}\rangle &=&\frac{\displaystyle 3-\sqrt{3}}{\displaystyle
6}|k\rangle|u_{ij}^{(2)}\rangle-\frac{\displaystyle
3+\sqrt{3}}{\displaystyle
6}|j\rangle|u_{ik}^{(2)}\rangle+\frac{\displaystyle
\sqrt{3}}{\displaystyle 3}|i\rangle|u_{jk}^{(2)}\rangle\nonumber\\
|h_{ijk}^{'}\rangle &=&\frac{\displaystyle 3-\sqrt{3}}{\displaystyle
6}|j\rangle|u_{ik}^{(2)}\rangle-\frac{\displaystyle
3+\sqrt{3}}{\displaystyle
6}|k\rangle|u_{ij}^{(2)}\rangle+\frac{\displaystyle
\sqrt{3}}{\displaystyle 3}|i\rangle|u_{jk}^{(2)}\rangle
\end{eqnarray}
\end{widetext}
for $i\neq j\neq k$. It is easy to see that $|g\rangle$'s form
orthogonal bases for subspace $S_4$, while $|h\rangle$'s form
orthogonal bases for subspace $S_5$. And there are $n(n+1)(n-1)/3$
$|g\rangle$'s and $n(n+1)(n-1)/3$ $|h\rangle$'s. Thus we can
rearrange the footnotes, and get $|g_i\rangle$ and $|h_i\rangle$,
where $i$ can change from $1$ to $n(n+1)(n-1)/3$. From the explicit
expressions of $|g_i\rangle$ and $|h_i\rangle$, we can easily get
\begin{equation}
\langle g_i|h_j\rangle=-\frac{1}{2}\delta_{ij},
\end{equation}
and $\{|g_i\rangle\}$ and $\{|h_i\rangle\}$ form Jordan bases for
$S_4$ and $S_5$. The two density operators that we want to
distinguish can now be expressed as
\begin{eqnarray}
\rho_1&=&\frac{2}{n^2(n+1)}\bigg(P_{S_0}+\sum_{i=1}^{i_0}|g_i\rangle\langle
g_i|\bigg),\nonumber\\
\rho_2&=&\frac{2}{n^2(n+1)}\bigg(P_{S_0}+\sum_{i=1}^{i_0}|h_i\rangle\langle
h_i|\bigg),
\end{eqnarray}
where $i_0=n(n+1)(n-1)$. $P_{S_0}$ is the projection onto the
subspace $S_0$, and
\begin{equation}
P_{S_0}=\sum_{i\leqslant j\leqslant
k=1}^{i_0}|u_{ijk}^{(2)}\rangle\langle u_{ijk}^{(2)}|.
\end{equation}

Now, let $T_i$ be the two-dimensional space spanned by the
nonorthogonal but linearly independent vectors $|g_i\rangle$ and
$|h_i\rangle$. The $T_i$ form a decomposition of subspace $S_6$ into
$n(n+1)(n-1)$ mutually perpendicular two-dimensional subspaces. Our
next problem is how to distinguish the two pure sates in every
subspace $T_i$.

\section{Derivation and implementation of the optimal unambiguous
discrimination}\label{sec4} We now want to unambiguously
discriminate between the subspaces $S_4$ and $S_5$ in $S_6$. First,
we distinguish the Jordan Basis states $|g_i\rangle$ and
$|h_i\rangle$ within their subspace $T_i$. In subspace $T_i$, the
apriori probabilities of $|g_i\rangle$ and $|h_i\rangle$ are
$\eta_1$ and $\eta_2$ respectively. Here we will use the method
mentioned in \cite{r16}. Let us define $|g_i^\bot\rangle$ and
$|h_i^\bot\rangle$ the reciprocal states of $|g_i\rangle$ and
$|h_i\rangle$ respectively, where
\begin{equation}
\langle g_i^\bot|h_i\rangle=\langle h_i^\bot|g_i\rangle=0
\end{equation}
and then
\begin{eqnarray}\label{eq5}
|g_i^\bot\rangle &=& \frac{\displaystyle |g_i\rangle-\langle
h_i|g_i\rangle|h_i\rangle}{\displaystyle \sqrt{1-|\langle
h_i|g_i\rangle|^2}}\nonumber\\
&=& \frac{\displaystyle 2}{\displaystyle
\sqrt{3}}|g_i\rangle+\frac{\displaystyle 1}{\displaystyle
\sqrt{3}}|h_i\rangle,\nonumber\\
|h_i^\bot\rangle &=& \frac{\displaystyle |h_i\rangle-\langle
g_i|h_i\rangle|g_i\rangle}{\displaystyle \sqrt{1-|\langle
g_i|h_i\rangle|^2}}\nonumber\\
&=& \frac{\displaystyle 2}{\displaystyle
\sqrt{3}}|h_i\rangle+\frac{\displaystyle 1}{\displaystyle
\sqrt{3}}|g_i\rangle.
\end{eqnarray}
Thus, $|g_i\rangle$ and $|h_i\rangle$ can be rewritten as
\begin{eqnarray}\label{eq2}
|g_i\rangle &=&\frac{\displaystyle \sqrt{3}}{\displaystyle
2}|g_i^\bot\rangle- \frac{\displaystyle 1}{\displaystyle
2}|h_i\rangle,\nonumber\\
|h_i\rangle &=&\qquad\qquad\quad\ |h_i\rangle.
\end{eqnarray}

\begin{figure}
\begin{center}
\epsfig{figure=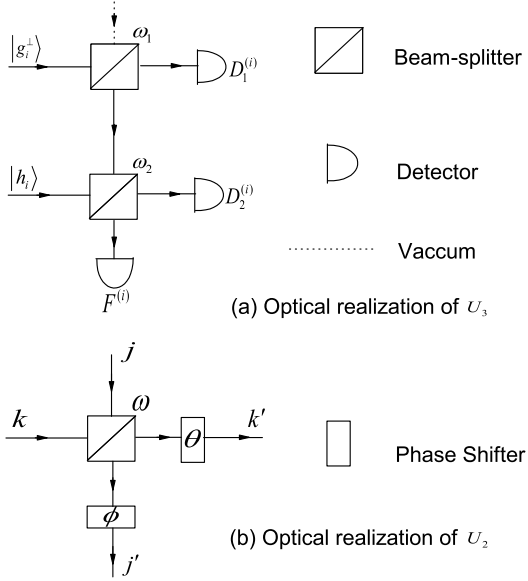,width=10cm}
\end{center}
\caption{(a) An optical six-port interferometer used to
unambiguously discriminate between $|g_i\rangle$ and $|h_i\rangle$.
(b) An optical four-port interferometer which is capable of
realizing the two-dimensional unitary transformation defined by Eq.
(\ref{eq6}).}
\end{figure}
We can give a physical implementation based on Neumark's theorem
\cite{r10}. Following the proposals given in \cite{r11,r12,r13,r14}:
(a) any pure state can be realized by a single-photon state and (b)
following Reck's theorem \cite{r14}, any unitary
transformation can also be realized by an optical network consisting
of beam-splitters, phase-shifter, etc. \cite{r15}, we can construct
an optical device, which is presented in Fig. 1a, to unambiguously
discriminate between $|g_i\rangle$ and $|h_i\rangle$. An additional
port is initially prepared in the vaccum state $|0\rangle$. The
operation of such a device is described by a unitary matrix $U_3$
which gives the probability amplitudes for a single photon entering
via inputs $|g_i^\bot\rangle$ and $|h_i\rangle$ to leave the device
by outputs $|D_1^{(i)}\rangle$, $|D_2^{(i)}\rangle$ and
$|F^{(i)}\rangle$. Here, the four-port optical interferometer, which
is presented in Fig. 1b, is used and it is capable of realizing any
$2\times 2$ unitary transformation $U_2$
\begin{equation}\label{eq6}
U_2=\left( \begin{array}{cc} \sin \omega e^{i\phi} & \cos \omega
e^{i\phi}\\
\cos \omega e^{i\theta} & -\sin \omega e^{i\theta}
\end{array} \right),
\end{equation}
where we have tacitly assumed that the action of all the $2\times 2$
beam splitter upon the states of the photon is described by a
unitary matrix of real coefficient
\begin{equation}
\left( \begin{array}{cc} \sin \omega & \cos \omega\\
\cos \omega & -\sin \omega
\end{array} \right).
\end{equation}
The parameter $\omega$ describes the transmittance ($\sqrt{t}=\sin
\omega$) and the reflectivity ($\sqrt{r}=\cos \omega$) of the beam
splitter. As shown in Fig. 1a, the action of this device, denoted by
$U_3$ gives,
\begin{eqnarray}
U_3|g_i^\bot\rangle &=& -\sin \omega_1|D_1^{(i)}\rangle+\cos
\omega_1\cos
\omega_2|D_2^{(i)}\rangle\nonumber \\
& & {}+ \cos \omega_1\sin \omega_2|F^{(i)}\rangle\nonumber\\
U_3|h_i\rangle &=& -\sin \omega_2|D_2^{(i)}\rangle+\cos
\omega_2|F^{(i)}\rangle
\end{eqnarray}
From Eq. (\ref{eq2}), $U_3$ will transform $|g_i\rangle$ and
$|h_i\rangle$ into
\begin{eqnarray}\label{eq3}
|g_i^{out}\rangle &=& U_3|g_i\rangle\nonumber\\
&=& -\frac{\displaystyle \sqrt{3}}{\displaystyle 2}\sin
\omega_1|D_1^{(i)}\rangle\nonumber\\
& &{}+(\frac{\displaystyle \sqrt{3}}{\displaystyle 2}\cos
\omega_1\cos \omega_2+\frac{\displaystyle 1}{\displaystyle 2}\sin
\omega_2)|D_2^{(i)}\rangle\nonumber\\
& &{}+(\frac{\displaystyle \sqrt{3}}{\displaystyle 2}\cos
\omega_1\sin \omega_2-\frac{\displaystyle 1}{\displaystyle 2}\cos
\omega_2)|F^{(i)}\rangle\\
\label{eq4}
|h_i^{out}\rangle &=& U_3|h_i\rangle\nonumber\\
&=& -\sin \omega_2|D_2^{(i)}\rangle+\cos \omega_2|F^{(i)}\rangle
\end{eqnarray}

As shown in Fig. 1a, both $|g_i\rangle$ and $|h_i\rangle$ have their
own detectors $D_1^{(i)}$ and $D_2^{(i)}$, and these two detectors
will tell us which the input is, while $F^{(i)}$ corresponds to
failure. This suggests that no photon can be detected by detector
$D_2^{(i)}$ $(D_1^{(i)})$ when the input is $|g_i\rangle$
$(|h_i\rangle)$. Thus, from Eq. (\ref{eq3}) we get
\begin{eqnarray}
\cos^2\omega_2 &=& \frac{\displaystyle 1}{\displaystyle
1+3\cos^2\omega_1},\nonumber\\
\sin^2\omega_2 &=& \frac{\displaystyle
3\cos^2\omega_1}{\displaystyle 1+3\cos^2\omega_1}.
\end{eqnarray}
Eq. (\ref{eq3}) and Eq. (\ref{eq4}) are reduced to
\begin{eqnarray}
|g_i^{out}\rangle &=& U_2|g_i\rangle\nonumber\\
&=& -\frac{\displaystyle \sqrt{3}}{\displaystyle 2}\sin
\omega_1|D_1^{(i)}\rangle\nonumber\\
& &{}+(\frac{\displaystyle \sqrt{3}}{\displaystyle 2}\cos
\omega_1\sin \omega_2-\frac{\displaystyle 1}{\displaystyle 2}\cos
\omega_2)|F^{(i)}\rangle,\\
|h_i^{out}\rangle &=& U_3|h_i\rangle\nonumber\\
&=& -\sqrt{\frac{\displaystyle 3\cos^2\omega_1}{\displaystyle
1+3\cos^2\omega_1}}|D_2^{(i)}\rangle\nonumber\\
& &{}+\sqrt{\frac{\displaystyle 1}{\displaystyle
1+3\cos^2\omega_1}}|F^{(i)}\rangle.
\end{eqnarray}

In other words, we can choose
\begin{eqnarray}
\Pi_1^{(i)out} &=& |D_1^{(i)}\rangle\langle D_1^{(i)}|,\nonumber\\
\Pi_2^{(i)out} &=& |D_2^{(i)}\rangle\langle D_2^{(i)}|,\nonumber\\
\Pi_0^{(i)out} &=& |F^{(i)}\rangle\langle F^{(i)}|,
\end{eqnarray}
with
\begin{eqnarray}
\Pi_1^{(i)out}|h_i^{out}\rangle=\Pi_2^{(i)out}|g_i^{out}\rangle &=& 0,\nonumber\\
\Pi_1^{(i)out}+\Pi_2^{(i)out}+\Pi_0^{(i)out} &=& \textbf{I},
\end{eqnarray}
to be our detection operators. Here, also from Fig. 1a, we can get
\begin{eqnarray}
|D_1^{(i)}\rangle &=& -\sin\omega_1|g_i^\bot\rangle,\nonumber\\
|D_2^{(i)}\rangle &=& -\frac{\displaystyle 2}{\displaystyle
\sqrt{3}}\sin\omega_2|h_i^\bot\rangle,
\end{eqnarray}
where Eq. (\ref{eq5}) has been used.

By projecting these detection operators back onto the space $T_i$,
we can get
\begin{eqnarray}
\Pi_1^{(i)} &=& \sin^2\omega_1|g_i^\bot\rangle\langle
g_i^\bot|,\nonumber\\
\Pi_2^{(i)} &=& \frac{\displaystyle 4}{\displaystyle
3}\sin^2\omega_2|h_i^\bot\rangle\langle h_i^\bot|\nonumber\\
&=& \frac{\displaystyle 4\cos^2\omega_1}{\displaystyle
1+3\cos^2\omega_1}|h_i^\bot\rangle\langle h_i^\bot|,\nonumber\\
\Pi_0^{(i)} &=& \textbf{I}-\Pi_1^{(i)}-\Pi_2^{(i)}
\end{eqnarray}
which form the POVM detection operators for the unambiguous
discrimination between $|g_i\rangle$ and $|h_i\rangle$.

So we find
\begin{eqnarray}
\langle g_i|\Pi_1^{(i)}|g_i\rangle &=& \sin^2\omega_1|\langle
g_i|g_i^\bot\rangle|^2\nonumber\\
&=& \frac{\displaystyle 3}{\displaystyle 4}\sin^2\omega_1,\nonumber\\
\langle h_i|\Pi_2^{(i)}|h_i\rangle &=& \frac{\displaystyle
4\cos^2\omega_1}{\displaystyle 1+3\cos^2\omega_1}|\langle
h_i|h_i^\bot\rangle|^2\nonumber\\
&=& \frac{\displaystyle 3\cos^2\omega_1}{\displaystyle
1+3\cos^2\omega_1}.
\end{eqnarray}
And the probability of successfully identifying the two states is
\begin{eqnarray}
P^{i}(\omega_1) &=& \eta_1\langle
g_i|\Pi_1^{(i)}|g_i\rangle+\eta_2\langle
h_i|\Pi_2^{(i)}|h_i\rangle\nonumber\\
&=& \frac{\displaystyle 3}{\displaystyle
4}\eta_1\sin^2\omega_1+\frac{\displaystyle
3\eta_2\cos^2\omega_1}{\displaystyle 1+3\cos^2\omega_1}.
\end{eqnarray}
By letting $x=1+3\cos^2\omega_1 (1\leqslant x\leqslant4)$,
$P^{i}(\omega_1)$ can be rewritten as
\begin{equation}
P^{i}(x)=1-\frac{\displaystyle \eta_1}{\displaystyle
4}x-\frac{\displaystyle \eta_2}{\displaystyle x}.
\end{equation}
This function, $P^{i}(x)$, has the property that $dP^{i}(x)/dx=0$
happens at
\begin{equation}
x=x_0=2\sqrt{\frac{\displaystyle \eta_2}{\displaystyle \eta_1}}.
\end{equation}
The optimal value of $P^{i}(x)$ denoted by ${P^{i}}^{opt}$ in the
domain $1\leqslant x\leqslant4$,
can be gotten in following three cases: \\
(a) if $\frac{\displaystyle 1}{\displaystyle 5}\leqslant
\eta_1\leqslant\frac{\displaystyle 4}{\displaystyle 5}$, the
requirement $1\leqslant x\leqslant4$ is satisfied, thus we have
\begin{equation}
{P^i}^{opt}=P^i(x=x_0)=1-2\sqrt{\eta_1\eta_2},
\end{equation}
(b) if $\eta_1<\frac{\displaystyle 1}{\displaystyle 5}$, we should
choose
\begin{equation}
{P^i}^{opt}=P^i(x=4)=\frac{\displaystyle 3}{\displaystyle 4}\eta_2,
\end{equation}
and (c) if $\eta_1>\frac{\displaystyle 4}{\displaystyle 5}$,
\begin{equation}
{P^i}^{opt}=P^i(x=1)=\frac{\displaystyle 3}{\displaystyle 4}\eta_1.
\end{equation}
Our finding can be summarized as follows
\begin{equation}
{P^i}^{opt}=\left\{ \begin{array}{ll} \frac{\displaystyle
3}{\displaystyle 4}\eta_2 & \textrm{if $\eta_1<\frac{\displaystyle
1}{\displaystyle 5}$}\\ \\
1-2\sqrt{\eta_1\eta_2} & \textrm{if $\frac{\displaystyle
1}{\displaystyle 5}\leqslant \eta_1\leqslant\frac{\displaystyle
4}{\displaystyle 5}$}\\ \\
\frac{\displaystyle 3}{\displaystyle
4}\eta_2 & \textrm{if $\eta_1>\frac{\displaystyle 4}{\displaystyle
5}$}
\end{array} \right..
\end{equation}

\begin{figure}
\begin{center}
\epsfig{figure=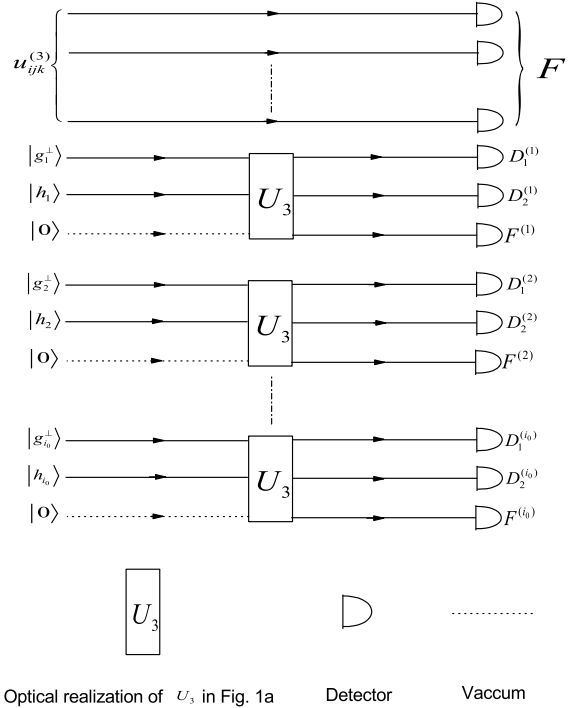,width=10cm}
\end{center}
\caption{Implementation of the programmable discriminator between
$\rho_1$ and $\rho_2$ ($|\Psi_1\rangle$ and $|\Psi_2\rangle$), which
consists of each implementation of discrimination between
$|g_i\rangle$ and $|h_i\rangle$.}
\end{figure}
Now, the discrimination between the two density operators $\rho_1$
and $\rho_2$ is to achieve $n(n+1)(n-1)/3$ discriminations described
above simultaneously. But, here the probability for the occurrence
of the input state in $T_i$ is $p(T_i)=\frac{\displaystyle
2}{\displaystyle n^2(n+1)}$. We can also give its implementation in
Fig. 2. We leave $|u_{ijk}^{(2)}\rangle$ alone, since they are in
both subspaces $S_1$ and $S_2$, and can't be distinguished. The POVM
which unambiguously distinguishes $S_1$ and $S_2$ has the form
\begin{eqnarray}
\Pi_1 &=&
\sum_{i=1}^{i_0}\Pi_1^{(i)}\nonumber\\
&=& \sin^2\omega_1\sum_{i=1}^{i_0}|g_i^\bot\rangle\langle
g_i^\bot|\nonumber,\\
\Pi_2 &=& \sum_{i=1}^{i_0}\Pi_2^{(i)}\nonumber\\
&=& \frac{\displaystyle 4\cos^2\omega_1}{\displaystyle
1+3\cos^2\omega_1}\sum_{i=1}^{i_0}|h_i^\bot\rangle\langle
h_i^\bot|,\nonumber\\
\Pi_0 &=& \textbf{I}-\Pi_1-\Pi_2,
\end{eqnarray}
where $i_0=n(n+1)(n-1)/3$. Thus, the probability of successfully
identifying the two density operators $\rho_1$ and $\rho_2$ is
\begin{eqnarray}
\widetilde{P}(\omega_1) &=&
\eta_1\textrm{Tr}(\Pi_1\rho_1)+\eta_2\textrm{Tr}(\Pi_2\rho_2)\nonumber\\
&=& \frac{\displaystyle 2\sin^2\omega_1}{\displaystyle
n^2(n+1)}\sum_{i=1}^{i_0}|\langle g_i|g_i^\bot\rangle|^2\nonumber\\
& &{}+\frac{\displaystyle 8\cos^2\omega_1}{\displaystyle
n^2(n+1)(1+3\cos^2\omega_1)}\sum_{i=1}^{i_0}|\langle
h_i|h_i^\bot\rangle|^2\nonumber\\
&=& \frac{\displaystyle (n-1)\eta_1}{\displaystyle
2n}\sin^2\omega_1+\frac{\displaystyle2(n-1)\eta_2\cos^2\omega_1}{\displaystyle
n(1+3\cos^2\omega_1)}.
\end{eqnarray}
Then, the optimal success probability $\widetilde{P}^{opt}$ can be
expressed as
\begin{equation}
\widetilde{P}^{opt}=\left\{ \begin{array}{ll} \frac{\displaystyle
n-1}{\displaystyle 2n}\eta_2
 & \textrm{if $\eta_1<\frac{\displaystyle 1}{\displaystyle 5}$}\\ \\
\frac{\displaystyle 2(n-1)}{\displaystyle 3n}(1-\sqrt{\eta_1\eta_2})
 & \textrm{if $\frac{\displaystyle 1}{\displaystyle 5}\leqslant \eta_1\leqslant
 \frac{\displaystyle 4}{\displaystyle 5}$}\\ \\
\frac{\displaystyle n-1}{\displaystyle 2n}\eta_1 & \textrm{if
$\eta_1>\frac{\displaystyle 4}{\displaystyle 5}$}
\end{array} \right..
\end{equation}

Now, we come back to the original problem that discrimination
between two pure states $|\psi_1\rangle$ ($|\Psi_1\rangle$) and
$|\psi_2\rangle$ ($|\Psi_2\rangle$). As mentioned before, the POVM
formed by $\Pi_1$, $\Pi_2$ and $\Pi_0$ with
\begin{equation}
\Pi_1|\Psi_2\rangle=\Pi_2|\Psi_1\rangle=0
\end{equation}
can be used. Its implementation is the same as discrimination
between $\rho_1$ and $\rho_2$ in Fig. 2. The success probability
is
\begin{eqnarray}
P(\omega_1) &=& \eta_1\langle\Psi_1|\Pi_1|\Psi_1\rangle+\eta_2\langle\Psi_2|\Pi_2|\Psi_2\rangle\nonumber\\
&=& (\frac{\displaystyle 1}{\displaystyle
2}\eta_1\sin^2\omega_1+\frac{\displaystyle
2\eta_2\cos^2\omega_1}{\displaystyle
1+3\cos^2\omega_1})(1-|\langle\psi_1|\psi_2\rangle|^2),\nonumber\\
\end{eqnarray}
where the relationship
\begin{eqnarray}
\sum_{i=1}^{i_0}\langle\Psi_1|g_i^\bot\rangle\langle
g_i^\bot|\Psi_1\rangle &=&
\sum_{i=1}^{i_0}\langle\Psi_2|h_i^\bot\rangle\langle
h_i^\bot|\Psi_2\rangle\nonumber\\
&=&\frac{\displaystyle 1}{\displaystyle
2}(1-|\langle\psi_1|\psi_2\rangle|^2)
\end{eqnarray}
has been used. Finally, we can also get the optimal success
probability as follows
\begin{equation}
P^{opt}=\left\{ \begin{array}{ll} \frac{\displaystyle
1}{\displaystyle 2} \eta_2(1-|\langle\psi_1|\psi_2\rangle|^2) &
\textrm{if $\eta_1<\frac{\displaystyle 1}{\displaystyle 5}$}\\
\\\frac{\displaystyle 2}{\displaystyle
3}(1-\sqrt{\eta_1\eta_2})(1-|\langle\psi_1|\psi_2\rangle|^2) &
\textrm{if $\frac{\displaystyle 1}{\displaystyle 5}\leqslant
\eta_1\leqslant
 \frac{\displaystyle 4}{\displaystyle 5}$}\\ \\
\frac{\displaystyle 1}{\displaystyle 2}
\eta_1(1-|\langle\psi_1|\psi_2\rangle|^2) & \textrm{if
$\eta_1>\frac{\displaystyle 4}{\displaystyle 5}$}
\end{array} \right.,
\end{equation}
which is apparently independent of $n$, the dimension of space
$\mathcal{H}$, as we mentioned at the beginning.

\section{Discussion and conclusion}\label{sec5}
\begin{figure}
\begin{center}
\epsfig{figure=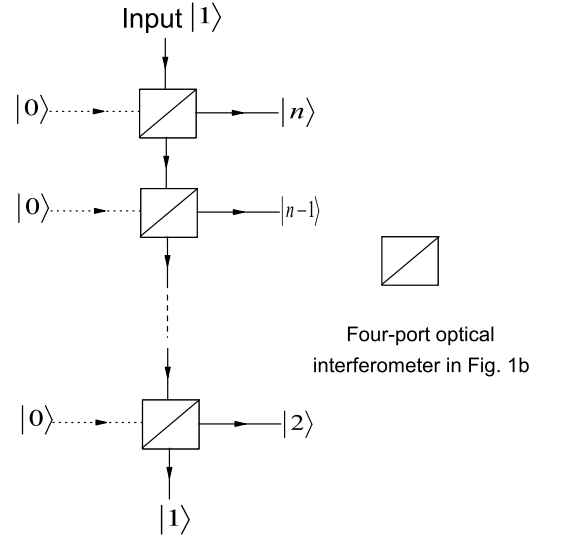,width=10cm}
\end{center}
\caption{An optical device to prepare the state
$|\psi\rangle=\sum_{i=1}^n|i\rangle$.}
\end{figure}
How to prepare the states with single photon is another important
question in the optical realization of the discriminator. We shall
give a brief discussion about this. An optical setting shown in Fig.
3 can prepare the state $|\psi\rangle=\sum_{i=1}^{n}a_i|i\rangle$.
It is constructed by a series of $2\times2$ unitary transformations
which can be realized by the four-port optical interferometer in
Fig. 1b. If the parameters of each four-port interferometer are
fixed properly, this setting achieves a single-photon state in
n-dimensional space $\mathcal{H}$. We denote the operation of this
device by $U_{|\psi\rangle}$, and
\begin{equation}
|\psi\rangle=U_{|\psi\rangle}|i\rangle.
\end{equation}
Then, by using $U_{|\psi_1\rangle}$, $U_{|\psi_?\rangle}$ and
$U_{|\psi_2\rangle}$ in succession, we can get
\begin{eqnarray}
|\Psi_?\rangle&=&|\psi_1\rangle\otimes|\psi_?\rangle\otimes|\psi_2\rangle\nonumber\\
&=& U_{|\psi_2\rangle}U_{|\psi_?\rangle}U_{|\psi_1\rangle}|1\rangle\nonumber\\
&=& U_{|\psi_1\rangle}\otimes U_{|\psi_?\rangle}\otimes
U_{|\psi_2\rangle}\cdot|1\rangle\otimes|1\rangle\otimes|1\rangle
\end{eqnarray}
where $|\psi_?\rangle$ is either $|\psi_1\rangle$ or
$|\psi_2\rangle$ and $|1\rangle\otimes|1\rangle\otimes|1\rangle$
denote the only input for the device. And $|\Psi_1\rangle$ or
$|\Psi_2\rangle$ is to be prepared depending on $U_{|\psi_?\rangle}$
being $U_{|\psi_1\rangle}$ or $U_{|\psi_2\rangle}$.

In conclusion, we have reconsidered the problem of the universal
programmable quantum state discriminator originally introduced in
\cite{r1}. And we solved a more generalized problem that the
unknown states are qudit states in the $n$-dimensional ($n\geqslant2$) 
Hilbert space $\mathcal{H}$.

We adopted the equivalence between the discrimination of unknown
pure quantum states and that of known mixed states. With the
Jordan-basis method, we simplified the problem and finally achieved
the optimal unambiguous discrimination between the two unknown
states, and the corresponding detection operators have already been
given. Significantly, we arrived at an important conclusion that the
optimal success probability of the discrimination between two
unknown pure states is independent of $n$, the dimension of the
space $\mathcal{H}$.

Furthermore, we also give the implementation of the optimal POVM
based on the Neumark's theorem. The POVM can be implemented on a
larger Hilbert space, where the additional degrees of freedom called
ancilla are needed.

\end{document}